\def\k{\kappa}
\def\f{\phi}
\def\l{\lambda}
\def\lb{\bar\l}
\def\p{\partial}
\def\ft{\phi^2}
\def\ff{\phi^4}
\def\e{\varepsilon}
\def\GN{\Gamma^{(N)}}
\def\gf{\gamma_{\f}}
\def\gft{\gamma_{\f^2}}

\def\i{\infty}
\def\frac#1#2{{{#1}\over{#2}}}
\def\s{\scriptstyle}

\def\ra{\rightarrow}
\def\disp{\displaystyle}
\def\b{\beta}
\def\L{\Lambda}
\def\G{\Gamma}
\def\y{m L}
\def\M{\cal M}

\def\t{\theta}
\def\r{\rightarrow}
\def\Gc{\cal G}
\def\F{\cal F}

\documentstyle[12pt]{article}

\makeatletter                                             

\def\section{\@startsection {section}{1}                  
{\z@}{-1.5ex plus -.5ex                                   
minus -.2ex}{1ex plus .2ex}{\large\bf}}                   

\def\@thmcountersep{}                                     

\long\def\@makecaption#1#2{\vskip 10pt                    
\setbox\@tempboxa\hbox{#1. #2}                            
  \ifdim \wd\@tempboxa >\hsize 
     #1. #2\par      
     \else                        
       \hbox to\hsize{\hfil\box\@tempboxa\hfil}           
   \fi}                                                   

\def\ps@headings{
 \def\@oddhead{\footnotesize\rm\hfill\runninghead\hfill}
 \def\@evenhead{\@oddhead}
 \def\@oddfoot{\rm\hfill\thepage\hfill}\def\@evenfoot{\@oddfoot}
}      

\makeatother

\title{Geometry, the Renormalization Group \linebreak and
Gravity}{}{}

\def\runninghead{O'CONNOR \& STEPHENS :\quad GEOMETRY, THE RG
AND GRAVITY}

\author{{\em Denjoe O'Connor} \thanks{D.~I.~A.~S., 10 Burlington Road,
Dublin 4, Ireland.}
\and
{\em C.~R.~Stephens} \thanks{Instituut voor Theor. Fysica,
Universiteit Utrecht,
3508TA Utrecht, Netherlands.}}

\date{} 

\begin{document}

\pagestyle{headings}
\flushbottom

\maketitle
\vspace{-10pt} 

\begin{abstract}

We discuss the relationship between geometry, the
renormalization group (RG) and gravity. We begin by reviewing
our recent work on crossover problems in field theory.
By crossover we mean the interpolation between different
representations of the
conformal group by the action of relevant operators. At the
level of the RG
this crossover is manifest in the flow between different fixed
points induced by these operators. The description of such flows
requires a RG which is capable of
interpolating between qualitatively different degrees of
freedom. Using the conceptual notion of course graining we
construct some simple examples of such  a group introducing the
concept of a ``floating''
fixed point around which one constructs a perturbation theory.
Our consideration of crossovers indicates that one should
consider classes of field theories, described by a set of
parameters, rather than focus on a particular one. The space of
parameters has a natural metric structure. We examine the
geometry of this space in some simple models and draw some
analogies between this space,
superspace and minisuperspace

\end{abstract}

\section{Introduction} 

The cosmopolitan nature of Charlie Misner's work is one of its
chief features.
It is with this in mind that we dedicate this article on the
occasion of his 60th birthday.
There are several recurring leitmotifs throughout theoretical
physics; prominent
amongst these would be geometry, symmetry, and fluctuations.
Geometry clarifies and systematizes
the relations between the quantities entering into a theory,
e.g. Riemannian geometry in the
theory of gravity and symplectic geometry in the case of
classical mechanics. Symmetry performs
a similar role, and in the case of continuous symmetries is
often intimately tied to geometrical
notions. For instance in the above examples Riemannian geometry
and symplectic geometry are intimately
related to the diffeomorphism and canonical groups respectively.
Our third leitmotif, fluctuations, enters
ubiquitously through the quantum principle, or classically in
statistical physics.
The key underlying idea here is that because of the fluctuations
physics must be described in a
probabilistic manner.

Having stated our prejudices let us be a little less ambitious
than to consider all of theoretical
physics and restrict our attention to field theory. We make no
pretension to mathematical rigour taking the point of view that
a field theory on a manifold $\cal M$ can be defined via a
functional integral with a probability measure which is a
functional of a set of possibly position dependent parameters
$\{g^i\}$, e.g. coupling constants, masses, background fields
etc. Physical quantities can be expressed as combinations of
moments which in turn can be written as functions of the
$\{g^i\}$. If we think of these parameters as coordinates on a
parameter space $\cal G$ it is clear that physics should be
invariant under changes in these coordinates. A particular type
of coordinate change is engendered by a renormalization, e.g.
between bare and renormalized $g^i$'s. Other possible symmetry
group transformations such as coordinate transformations on
$\cal M$ or gauge transformations act as diffeomorphisms on
$\cal G$. Here we are concerned   exclusively with the behaviour
under RG transformations, and hence under scale transformations.
We investigate some geometrical structures on $\cal G$, in
particular defining a metric and associated connection. We look
at the change in the geometry under renormalization, thereby
introducing all three of our leitmotifs. The geometry is a
result of the fluctuations in the system, i.e the probabilistic
description. Without fluctuations the metric is identically
zero. The RG induces a flow on $\cal G$ the fixed points of
which are of particular interest as they represent conformally
invariant systems. This flow with respect to a given parameter
can be either centrifugal or centripetal for a particular fixed
point. If the former the parameter is said to be relevant, and
irrelevant for the latter. The relevance or irrelevance can
change according to the fixed point.

RG flows between different fixed points, i.e different conformal
field theories, are especially interesting. The reason for this
is the following: one of the most important tasks confronting a
theory is to identify correctly the degrees of freedom (DOF) of
a physical system. It is a fact of life that all physically
relevant theories have qualitatively different effective DOF at
different scales. For instance, in QCD the high energy DOF are
quark, gluon DOF, whilst at low energy they are hadron, meson
... DOF. In gravity at low energy, gravitons are the low energy
DOF, whereas at high energies, who knows...topological foam,
strings .... The only thing that is reasonably certain is that
it won't be gravitons. A closer to earth example would be liquid
helium in a 3 dimensional (3D) slab geometry. For correlation
lengths much less than the slab thickness helium atoms are the
relevant DOF whereas in the opposite limit it is vortices. An
example we will treat here is that of a $\l\ff$ theory on a
manifold $S^1\times R^{d-1}$ of size $L$. Suitably altered this
model model can describe,
amongst others, the Higgs model at finite temperature, the
Casimir effect for an interacting quantum field theory or
the critical behaviour of an Ising ferromagnet in a slab
geometry. Here there is a change in DOF as the variable $x=mL$
changes, where $m$ is the ``mass'' (inverse correlation length)
in the physical system. As $x\ra\i$ the DOF are effectively $d$
dimensional and as $x\ra0$, $d-1$ dimensional. We will also
briefly discuss similar considerations in more realistic
``cosmologies''.

One of the first questions one must confront with a crossover
problem is: how should one renormalize? If one accepts the
fairly common point of view that renormalization means the
consistent removal of ultraviolet (UV) divergences one
generically finds a resultant RG which is independent of the
parameter inducing the crossover, e.g. $L$ in the above example.
The $\b$ functions and anomalous dimensions of the problem are
all then $L$ independent. One also finds that the theory gives
perturbative nonsense as $x\ra0$. The reason for this is
relatively simple. Let us take a more physical picture of
renormalization, as a ``course graining'' such as
decimation/block spinning\footnote{Strictly speaking such
renormalizations form a semigroup not a group.} [1]. Here we
imagine integrating out DOF between one scale and another. For
the finite system at scales $\ll L$ one would integrate out $d$
dimensional DOF. However, as one course grains further one is
eventually integrating out DOF with scales $\sim L$. In the
finite direction there are no DOF with scales $>L$, therefore
one cannot integrate them out. The only DOF left are $d-1$
dimensional and these are the physically relevant ones. So, a
physically intuitive renormalization procedure takes into
account the qualitatively changing nature of the effective DOF.
It should be clear then why a $L$ independent RG is badly
behaved. Such a group is equivalent to integrating out only $d$
dimensional DOF {\bf for all scales}. The moral is that one
should try to develop a RG that is capable of interpolating
between qualitatively different DOF. In this paper we will show
how this can be achieved in a wide class of crossover problems.

The outline of this article will be as follows: in section 2 we
will give a short, intuitive exposition of renormalization and
the RG with a view to the treatment of crossovers. In section 3
we will develop the concept of a RG that can interpolate between
qualitatively different DOF, introducing the concept of a
``floating'' fixed point and illustrating our ideas with $\l\ff$
on $S^1\times R^{d-1}$. In section 4 we will describe the
beginnings of a geometrical framework for field theory wherein a
much more general theory of crossovers may be built illustrating
the concepts using a Gaussian model. Finally in section 5 we
take an opportunity to make some speculative remarks and draw
some conclusions.

\section{Renormalization and the Renormalization Group} 

In this section as well as setting notation we would like to
give an extremely brief and hopefully intuitive account of
renormalization, hoping that the unconventional viewpoint will
not prove unintelligible. As a concrete example consider a self
interacting scalar field theory described by a partition
function (generating functional) on ${\cal M}=R^d$
$$Z[m_B,\l_B,\L]=\int[D\f_B]_{\L}e^{-{\int_{\L} d^dx
L(\f_B,m_B,\l_B)}}\eqno(1)$$
where
$$L={1\over2}(\nabla\phi)^2+{1\over2}m^2_B\phi^2_B+
{\lambda_B\over4!}\L^{4-d}\phi^4_B
\eqno(2)$$
For the sake of making sense out of the theory we will assume
there is always
an UV cutoff $\L$.
In (1) we have a probability measure which is a function of two
parameters and a cutoff. The parameters $m_B$ and $\l_B$ are
good descriptors of the physics at scales $\sim\L$. What this
means is the
following; if one could calculate the 2 and 4 point vertex
functions exactly one would find them to be very complicated
functions of $m_B$, $\l_B$ and $\L$. At scales $\sim\L$,
however, one would
find that for $\l_B<<1$ $\G^{(2)}\sim m_B$ and $\G^{(4)}\sim
\l_B$. On the other hand at scales $\k<<\L$ the parameters $m_B$
and $\l_B$ are in no way a good description of the associated
correlation functions. Obviously as
${\L\over\k}\ra\i$ they get worse and worse. The deep underlying
reason behind this is of course the existence of fluctuations.
It is the dressing due to quantum or thermal fluctuations that
changes the correlation functions as one changes scale. We
emphasize though that if one can calculate in the theory exactly
the bare parameters are as good as any others.
What one would like is to describe the correlation functions
using a more suitable set of parameters, in particular if we are
considering physics at a
scale $\sim\k$ it would seem to make good sense to describe the
physics using
new parameters $m$ and $\l$ which are a more natural description
of the physics at this scale. An obvious natural choice would be
to describe the physics at the scale $\k$ in terms of the 2 and
4 point vertex functions at a scale $\k'$ where $\k\sim\k'$.
Thus one would require
$$\G^{(2)}(k=0,m,\l,\k')=m^2 \; \; \; \; \; \; \;
\G^{(4)}(k=0,m,\l,\k')=\lb=\l\k'^{4-d} \eqno(3)$$
The physics at the scale $\k$, i.e the correlation functions at
that scale, would now be described in terms of the correlation
functions at a nearby fiducial scale, $\k'$.

In the above we have loosely outlined the renormalization
program for this model. Why renormalize? There are two answers
to this, one perturbative, and one not. Perturbatively as
${\L\over\k}$ grows perturbation theory in terms of the bare
coupling becomes worse and worse. This is the well known problem
of ``UV divergences''. In terms of fluctuations the bare
parameters are being perturbatively dressed by fluctuations
between the scales $\L$ and $\k$. The recipe for getting round
this problem is as outlined above; to perturb with a ``small''
coupling rather than
a large one, i.e. the renormalized coupling. Thus one uses the
value of $\G^{(4)}$ at some scale $\k'$ as one's small
parameter. This perturbation theory is then reasonably well
defined as long as $\k$ is not too different from $\k'$ as. In
4D, for example, the correction terms are proportional to powers
of ${\rm ln}{\k\over\k'}$. Thus it is perturbatively better to
dress the correlation functions a small amount. The dressing
between $\L$ and $\k$ is large and therefore difficult to
compute perturbatively whilst the dressing between $\k$ and
$\k'$ is smaller. The optimum approach is to consider an
infinitesimal dressing and to integrate the resulting
differential equation.
So, if one wishes to implement perturbation theory
renormalization is essential.
The non-perturbative reason is somewhat subtler and depends
ultimately on whether one believes there is a fundamental cutoff
or not. One puts it in to make mathematical sense of the theory
and then asks if it can be sensibly removed again. It seems to
be the case that this is only possible for special values of the
bare parameters --- their fixed point values. To understand this
we must consider the RG.

One can think of renormalization as a mapping of correlation
functions between two different ``scales''. These mappings have
an abelian group structure and this group is known as the RG.
The group action on $\cal G$ generates a flow. Of particular
interest are the fixed points of this flow as they imply a
system possesses scale invariance. The fundamental relation
between bare and renormalized vertex functions is
$$\GN(k,m,\l,\k)=Z^{N\over2}_{\f}\GN_B(k,m_B,\l_B,\L)\eqno(4)$$
where the renormalized parameters are defined at some arbitrary
scale $\k$, and $Z_{\f}$ denotes a wavefunction renormalization
factor. The bare theory's independence from $\k$ leads to the RG
equation
$$\left(\kappa{\partial\over\partial\kappa}
+\beta{\partial\over\partial\lambda}
+\gamma_{\phi^{2}}m^2{\partial\over\partial m^2}
-{N\over2}\gamma_\phi\right)
\Gamma^{(N)}(k_i,m^2,\lambda,\kappa)=0
\eqno(5)$$
where $\gft=-{\p{\rm ln}Z_{\ft}\over\p{\rm ln}\k}$, $Z_{\ft}$
being the renormalization constant associated with the operator
$\ft$ and $\gf={\p{\rm ln}Z_{\f}\over\p{\rm ln}\k}$ are the
anomalous dimensions of the operators $\ft$ and $\f$
respectively. It is important to note that $(5)$ results from an
{\bf exact} symmetry even though it expresses an apparent
triviality, the reparameterization invariance of the correlation
functions. Equation $(5)$ can be solved by the method of
characteristics and together with dimensional analysis yields
$$\Gamma^{(N)}(k_i,\lambda,m,\kappa)
=(\kappa\rho)^{d-N{(d-2)\over 2}}{\rm exp}\left({N\over 2}
\int^{1}_{\rho}{dx\over x}\gamma_{\phi}(x)\right)
\Gamma^{(N)}({k_i\over\rho\kappa},
{m^2(\rho)\over\rho^2\kappa^2},\lambda(\rho),1)
\eqno(6)$$
where $\lambda(1)=\lambda$, $m(1)=m$, and $\rho$ is arbitrary.
$m(\rho)$ and $\l(\rho)$, the running mass and coupling satisfy
$$
\rho{d\l(\rho)\over{d\rho}}=\beta
\; \; \; \; \rho{dm^2(\rho)\over{d\rho}}=\gft m^2(\rho)
\eqno(7)$$
Equation $(5)$ tells us how $\GN$ gets dressed by fluctuations
between the scales $\k$ and $\k+d\k$, in terms of parameters
which get dressed according to $(7)$. Integrating this equation
tells us how $\GN$ gets dressed by fluctuations between the
scales $\k$ and $\k\rho$. This dressing induces a flow on $\cal
G$. Equation $(6)$ is the exact solution of an exact equation
which is a result of an exact symmetry. The fixed point of the
coupling $\l$, $\l^*$ is given by the solution of $\b=0$. Now,
we can use our freedom in choosing $\rho$ to eliminate the
variable $m(\rho)$ in $(6)$ via the condition
$m^2(\rho)=\rho^2\k^2$. At the fixed point $\l^*$ one can solve
the equation for $m(\rho)$ to find
$\rho\sim{({m\over\k})}^{\nu}$ where $\nu=(2-\gft^*)^{-1}$,
$\gft^*$ being the value of $\gft$ at the fixed point.
Similarly, defining $\eta=\gf^*$ one finds for instance
$$\G^{(2)}(k=0,m)\sim Am^{2\nu(2-\eta)}\eqno(8)$$
where $A$ is some constant. Once again we emphasize that this is
an exact result dependent only on the fact that a fixed point
exists. The RG is not just about ``improving perturbation
theory''. Of course, finding the fixed point and calculating
$A$, $\nu$ and $\eta$ is a different matter. In $d<4$ dimensions
for this model there are two known fixed points, the Gaussian
fixed point $\l^*=0$ and the Wilson-Fisher (WF) fixed point
$\l^*\sim (4-d)$. At the Gaussian fixed point $\nu={1\over2}$
and $\eta=0$ whilst at the other e.g. in 3D $\nu=0.630$ and
$\eta=0.031$.

Physically the importance of the fixed point for $\l$ is the
following. $\l$ like all other quantities gets dressed as a
function of scale and therefore changes its value. At the point
$\l=\l^*$ the coupling becomes completely insensitive to
dressing and therefore has essentially dropped out of the
problem. Obviously $m=0$ is a fixed point for the mass. As fixed
points essentially define a theory finding them is one of the
main tasks of field theory. Returning now to a non-perturbative
aspect of renormalization; in $(4)$ we could instead of
differentiating the bare vertex function with respect to $\k$
have differentiated the renormalized function with respect to
$\L$. This yields an equation analogous to $(5)$. If one can
find a fixed point of this equation then one can take the cutoff
$\L\ra\i$ and thereby recover a continuum theory.

The fact that there exist two fixed points for this theory means
that one is really considering a class of field theories as a
function of $x={\lb\over m^{4-d}}$. As $\lb\ra0$ one approaches
the Gaussian theory, and as $m\ra0$ the WF fixed point. One
crosses between them as a function of scale. The coupling $\lb$
is relevant in terms of RG flows with respect to the Gaussian
fixed point. In other words a small perturbation from this fixed
point induces a flow to larger length scales terminating at the
WF fixed point. This is an example of crossover behaviour in
field theory and describes a transition between qualitatively
different DOF. For $x\ll1$ the DOF are essentially
non-interacting, whereas for $x\gg1$ they are strongly
interacting. The reader might legitimately enquire as to why,
given that they are strongly interacting, one believes that
perturbation theory can be used. This raises an important
question: perturbation theory in terms of what coupling? In
terms of $\lb$ straight perturbation theory breaks down as
$m\ra0$ due to large dressings from the infrared (IR) regime as
opposed to large dressings from the UV regime as was considered
previously. The RG methodology tells you to ignore any
differences between the UV and IR regimes. The essential problem
is that of large dressings irrespective of whether the dressing
arises from IR or UV fluctuations. Large dressings imply that
one has used inadequate parameters to describe the physics,
hence renormalization and the RG should be implemented. The
correct parameter to perturb in is the running coupling constant
which is a solution of the $\b$ function equation treated as a
differential equation who's solution is a function of $x$.
The above is our first simple example of crossover behaviour in
field theory. We would now like to proceed to other more
pertinent examples showing some difficulties one encounters and
their solution.

\section{Crossover Behaviour in Field Theory} 

One of the main themes we have tried to emphasize in the
introduction is that the effective degrees
of freedom of a physical system are scale dependent. Here we
take a
simple but physically relevant paradigm to show the difficulties
involved in trying to describe
a qualitative change in the DOF of a system. We will try to
emphasize a physical approach,
stating in general only results, leaving the details in our
other papers [2].
Consider a Lagrangian
$${\cal L}={1\over2}(\nabla\phi_B)^2+{\frac12}m^2_B\phi_B^2+
\sum_i\mu_B^iO_B^i\eqno(9)$$
where $\sum_i\mu_B^iO_B^i$ represents schematically a relevant
or set of relevant operators that induce a crossover from a
fixed point associated with $\mu^i=0$ to some others. For the
moment we specify neither the symmetry of the order parameter or
the dimensionality of the system. Some examples  of relevant
operators are the following: i) for $d<4$, Gaussian$\ra$WF fixed
point as mentioned in the last section, $\mu_B^1={\l_B\over4!}$,
$O^1=\ff_B$, $\mu^i_B (i\neq1)=0$; ii) quadratic symmetry
breaking ($O(N)\ra O(M)$), $\f_B$ has an $O(N)$ symmetry,
$\mu_B^1={\frac12}\tau_B$, $O_B^1=\sum_{i=1}^M(\f_{iB})^2$,
$\mu_B^2={\l_B\over4!}$, $O_B^2=\sum_{i=1}^N(\ft_{iB})^2$; iii)
uniaxial dipolar ferromagnets, where in Fourier space
$\mu_B^1={\alpha\over2}$, $O_B^1={p_z^2\over p^2}\f_B^2$,
$\mu_B^2={\l_B\over4!}$, $O_B^2=\ff_B$. For the case of
dimensional crossover one can determine the appropriate
operators by Fourier transforming $\cal L$ with respect to the
finite directions.
One important common feature of the above is the introduction of
an important new scale in each problem i.e. $\tau$, the
quadratic symmetry breaking term, $\alpha$ the strength of the
dipole-dipole interactions and $L$ the characteristic finite
size scale. It is the existence of one or more new scales in a
problem that makes a crossover much richer, more interesting and
more complex than standard field theory. We call this generic
length scale $g$. We also take this scale to be a physical scale
and hence a RG invariant.

So, what does renormalization have to say about such systems?
There is a widespread belief that renormalization just means
getting rid of UV divergences. If we accept this belief and
examine the above models one notices that the UV behaviour in
these theories is independent of the parameter $g$, hence the UV
divergences can be removed in a $g$ independent way. We will
give just one example of what happens if this philosophy is
accepted. Consider $\l\ff$ on  a manifold $S^1\times R^3$ of
size $L$. Using minimal subtraction gives for $mL\ll1$ to one
loop
$$\G^{(2)}\rightarrow m^2\left(1+{\l\over{32\pi^2}}{\rm
ln}{m^2\over{\k^2}}+{\l\over{24m^2L^2}}+O({\l^2\over
m^3L^3})\right)\eqno(10)$$
Obviously the perturbative corrections are large in this regime,
in fact in the limit $Lm\ra0$ these corrections become infinite.
{}From the point of view of renormalization this is no different
than the bare vertices in the $L$ independent theory getting a
large dressing due to fluctuations. Here we've done a
renormalization but still the vertex has a large $L$ dependent
dressing. Why is that? In implementing minimal subtraction we
have really made an assumption, that parameters associated with
the $L=\i$ system will provide a  good description of the
physics when $L\ra0$. In this limit the system is effectively 3D
and so one can hardly expect 4D parameters to be adequate. The
total breakdown in perturbation theory above is a reflection of
this fact. 3D $\l\ff$ theory has completely different DOF to 4D
$\l\ff$ theory.

The way out of this impasse is in many ways relatively simple
--- choose better renormalized parameters. Think back to the
G-WF crossover discussed in the last section. The analog of the
4D theory there is the Gaussian theory and the analog of the 3D
theory the WF fixed point theory. The analog of $\lb$ is $L$.
$L$ is a relevant parameter that causes a crossover from  one
fixed point to another. We managed to cope with the G-WF
crossover, how so? Above we renormalized in an $L$ independent
way, the analog would be to renormalize in a $\lb$ independent
manner. We could have certainly done this i.e renormalize the
theory using only the counterterms appropriate for a Gaussian
theory. For ${\L\over\k}\gg1$ we would have found large
dressings telling us
that the Gaussian counterterms were not really sufficient to
renormalize the theory. These large dressings occur because of
the self interactions amongst the particles, because interacting
DOF are qualitatively different to non-interacting ones. The
correct thing to do was to choose renormalization conditions
such as in (3) which were specified as functions of $\l$, i.e a
good renormalization was dependent on $\l$ the parameter that
induces the crossover. In the case at hand we should therefore
consider $L$ dependent renormalization conditions such as
$$\G^{(2)}(k=0,m,\l,L,\k)=m^2 \; \; \; \; \; \; \;
\G^{(4)}(k=0,m,\l,L,\k)=\l\k^{4-d} \eqno(11)$$
These conditions imply that the $\b$ function and anomalous
dimensions are all functions of $L\k$ as well as $\l$, i.e the
RG itself is $L$ dependent. An $L$ independent RG tells you how
parameters are dressed in the theory by $L$ independent
fluctuations whereas an $L$ dependent one tells how things are
dressed by $L$ dependent ones. In the real physical system it is
manifestly obvious that the real fluctuations in the system are
$L$ dependent and that consequently conditions such as $(11)$
will yield parameters which are a more faithful representation
of the physics. The moral is: if the DOF of a system can
qualitatively change as a function of scale then it is clearly
better if one can derive a RG which can follow such a change.

It should be clear how to implement this philosophy more
generally. For a crossover caused by a relevant parameter $g$,
one should impose normalization conditions at an arbitrary value
of $g$ thereby obtaining a $g$ dependent RG equation. In such a
crossover one is interpolating between different conformal field
theories i.e. different representations of the conformal groups
associated with the limits ${g\over m}\ra0$ and ${g\over
m}\ra\i$. Just as there are anomalous dimensions $\gf$ and
$\gft$ which are characteristic of the conformal weights of the
associated operators for a particular fixed point so one can
define effective anomalous dimensions and critical exponents
which are characteristic of the crossover system. Given that in
$d$ dimensions the dimension of the operator $\ff$ is
canonically $4-d$ one can define an effective dimension
$d_{eff}$ via the relation ${d{\rm ln}\G^{(4)}\over d{\rm
ln}m^2}=4-d_{eff}-2\eta_{eff}$. What about the notion of a fixed
point? For the system of size $L$ true conformal symmetry is
only realized in the limits $m\ra0$ and $Lm\ra\i$ which yields
the $d$ dimensional fixed point and $m\ra0$ $Lm\ra0$ which
yields the $d-1$ dimensional one. The equation $\b=0$ as an
algebraic equation still has some meaning in these crossover
systems. It does not, however, give a fixed point because the
$\b$ function is now explicitly scale dependent through the
variable $L\k$. If one thinks of the $\b$ function as being the
velocity of the RG flow in the $\l$ direction the value $\b=0$
is an equation that is satisfied only for a particular scale,
not all scales as it should be for a true fixed point. The $\b$
function equation is a differential equation and should be
integrated. However, one can in fact define an effective or
``floating'' fixed point in the following manner. Consider the
$\b$ function generically as
$$\k{d\l\over
d\k}=\b(\l,L\k)=-(4-d)\l+a_1(L\k)\l^2+a_2(L\k)\l^3+O(\l^4)
\eqno(12)$$
where $a_1$ and $a_2$ are known functions (see [3]). Define a
new coupling $h=a_1\l$ to find
$$\k{dh\over d\k}=-\e(L\k)h+h^2+b(L\k)h^3+O(h^4)\eqno(13)$$
where $\e(L\k)=4-d-{d{\rm ln}a_1\over d\ln\k}$ and $b(L\k)$ is a
combination of $a_2$ and $a_1$. Setting $\b(h,L\k)=0$ yields a
solution $h^*\equiv h^*(L\k)$. This is the floating fixed point.
As $L\k\ra\i$ it yields the $d$ dimensional fixed point and as
$L\k\ra0$ the $d-1$ dimensional fixed point. Corresponding
floating fixed points can be defined in all the crossover
problems we have considered so far. The floating fixed point is
the ``small'' parameter with which perturbation theory is
implemented. A $g$ dependent RG and a corresponding $g$
dependent RG improved perturbation theory allow for complete
perturbative access to the crossover. The main reason for this
is that such a RG can interpolate between the qualitatively
different DOF in the problem.
As a specific example we will quote some one loop results [2]
for the above finite size model. The fixed point coupling is
$h^*=\e(\k L)$ where
$$4-d_{eff}=\e(\k L)=1-\k{d\over d\k}{\rm ln}
({\disp\sum_n}(1+{4\pi^2n^2\over L^2\k^2})^{-\s{\frac32}})$$
$$\gft(h^*)={h^*\over3} \; \; \; \; \; \; \; \gf(h^*)=0$$
These functions all interpolate in a smooth way between their 4D
and 3D values. $\e(\k L)$ is our ``small'' expansion parameter.
It also yields (to first order) the effective dimensionality of
the system. It is worth noting here that the sole requirement of
finiteness of the correlation functions for all $L$ is
sufficient to determine the entire crossover.

So far we have outlined intuitively an approach to crossover
behaviour and applied it to an interesting class of problems.
Our considerations were governed by the RG flows of the
parameters. The natural arena for such flows is $\cal G$. Rather
than consider a particular crossover we would like to consider
$\cal G$ more abstractly. This may also prove fruitful in cases
where the relevant parameters are not a priori known.

\section{Geometry of $\cal G$} 

\def\f{\hat\phi}

In this section we wish to begin an investigation of some of the
geometrical
structure that is inherent in the approach we are following. We
will attempt
to be as general as possible to begin with, and consequently
somewhat vague.
As was seen in the preceeding sections it was essential, if one
wished to have a controlled perturbative expansion, to change
from one set of parameters useful in one regime to a new set of
parameters useful in another regime, for example the large $\y$
and small $\y$ regimes respectively. We are therefore working on
a coordinate patch and changing coordinates on this patch. The
immediate question would appear to be what space are we working
on, i.e. a coordinate patch of what?

Examining the functional integral
$$Z[{\M},\{\t^i\} ,\L]=\int[D\phi]_{\L}e^{-S[{\M},\phi[{\M
}],\{\t^i\},\L]}\eqno(14)$$
we see that it defines a map from the space, ${\F}$,
parameterized by
$({\M},\phi[{\M }],\{\t^i\},\L)$ to a section of a line bundle
over ${\Gc}$. ${\M}$ is the spacetime manifold, $\phi[{\M}]$ are
fields on ${\M}$, $\{\t^i\}$ are couplings  between the fields
and external sources and
$\L$ plays the role of a regulator which will not be viewed  as
a true parameter of the theory, but rather as either a
reflection of a true underlying lattice or a device to  control
any UV problems, and assist in the definition of the functional
integral. We choose the set $\{g^i\}$ discussed in previous
sections to be local coordinates on $\cal G$.

Earlier we saw that explicit calculations required a change of
parameters, i.e. a change of coordinates on $\cal G$, from bare
parameters (coordinates) to renormalized parameters
(coordinates). If the object $Z$ has any meaning it should have
the same content in all coordinate systems. We will assume that
$Z$ is invariant under coordinate transformations on ${\Gc}$ and
therefore is a scalar. Now, when one is looking at coordinate
transformations, it is natural to examine what structure ${\Gc}$
posesses that can help one organize the analysis. Any structure
${\Gc}$ has must be induced from $Z$, or already exist in $S$.
Ideally we would like our parameters to be related as simply as
possible to the moments of the probability distribution, as
these are generally the experimentally accessible objects.  We
will assume that ${\Gc}$ is a topological space with a
differentiable structure and possibly isolated singularities,
and  that $Z$ can be considered a differentiable function on
${\Gc}$ away from such special points. Thus if we consider an
infinitesimal variation in $S$ of the form $dS$, where $d$ is
the exterior derivative operator on ${\Gc}$, we get an induced
change in $Z$. If the sources, masses etc. are position
dependent then ${\Gc}$ is infinite dimensional and analogous to
superspace, which would suggest that a mini-superspace may be
useful. Mini-superspace in this context means restricting our
considerations to
a small finite dimensional subspace of ${\Gc}$. It is primarily
this situation that will concern us here.

It is convenient for the following to work with the functional
integral as a normalized probability distribution, which we can
achieve by dividing by $Z=e^{-W}$.
We therefore get a normalized functional integral
$$\int[D\phi]_{\L}e^{W-S} =1\eqno(15)$$
Because it is  normalized  and $d$ is restricted to ${\Gc}$,
we have
$$\int[D\phi]_{\L}de^{W-S}=dW-<dS>=0\eqno(16)$$
where $<A>$ means expectation value of $A$.
$$ds^2=<(dW-dS)\otimes(dW-dS)>\eqno(17)$$
defines a positive definite, symmetric, quadratic form on
${\Gc}$ arising from the positivity of the probability
distribution or the convexity of the associated entropy
functional.  $ds^2$ plays the role of a metric on ${\Gc}$.
An infinitesimal change in our parameters along some smooth
curve in ${\Gc}$ defines a vector tangent to that  curve  and
therefore we can express our metric as
$$g_{\mu\nu}= <\p_{\mu}S\p_{\nu}S>-\p_{\mu}W\p_{\nu}W\eqno(18)$$
on the space satisfying $dW-<dS>=0$. This metric is known as the
Fisher information matrix [4] in probability theory and is used
for comparing one probability distribution to another.

Let us begin with our most simple mini-superspace example, the
Gaussian distribution, which corresponds to a free field theory
in zero dimensions. We begin with a field $\phi$ coupled to an
external source $J$ described by the action
$$S[\phi,m^2,J]={1\over 2}m^2\phi^2+J\phi\eqno(19)$$
${\M}$ is now a single point and we have a coordinate patch on
${\Gc}$ with coordinates $(J,m^2)$. The generator of connected
correlation functions is
$$W[J,m^2]=-{1\over2}{J^2\over m^{2}}+{1\over2}{\rm
ln}[{m^2\over 2\pi}]\eqno(20)$$
The condition $dW-<dS>=0$ gives
$$-(<\phi>+{J\over m^2})dJ-{1\over2}(<\phi^2>-{J^2\over
m^4}-{1\over m^2})dm^2=0\eqno(21)$$
The corresponding metric on using (21) is
$$ds^2={1\over m^2}dJ^2 -{2\over m^2}{J\over m^2}dJ
dm^2+{1\over m^2}(({J\over m^2})^2+{1\over2}{1\over m^2})(dm^2)^2\eqno(22)$$
Note that this metric is not diagonal unless $J=0$, however, a
simple coordinate
change allows us to diagonalize it, the appropriate choice of
new coordinate being $\f=-{J\over m^2}$ which is equivalent to
starting with
$$S[\phi,\f,m^2]={1\over 2}m^2\phi^2-m^2\f\phi\eqno(23)$$
$$W[\f,m^2]=-{1\over2}m^2\f^2+{1\over2}{\rm ln}[{m^2\over
2\pi}]\eqno(24)$$
the condition $dW-<dS>=0$ now gives
$$m^2(<\phi>-\f)d\f-{1\over 2}(<(\phi-\f)^2>-{1\over
m^2})dm^2=0\eqno(25)$$
with  metric
$$ds^2=m^2d\f^2+{1\over 2}m^{-4}(dm^2)^2\eqno(26)$$
Observe that if $m^2$ were negative this metric would not be
positive definite and if $m^2=0$ it would be highly singular.
This is connected to stability,  unitarity and convexity of $W$.
It is not difficult to verify that this metric (in either
coordinate system) has scalar curvature $R=-{1\over 2}$. Before
discussing the meaning of this let us see what happens in a more
realistic field theoretic setting.

Consider a Gaussian model on a compact manifold ${\M}$ of volume
$L^d$, where $d\leq 4$ and
$$S[\phi,J,m^2,L,\L]=\int_{\M}
       [{1\over2}\phi(\Box+m^2)\phi+J\phi]\eqno(27)$$
$J$ and $m^2$ can be position dependent, and in fact generically
are on a curved ${\M}$. A coordinate transformation equivalent
to above gives
$$S[\phi,\f,m^2,L,\L]
=\int_{\M}[{1\over2}\phi(\Box+m^2)\phi-\phi({\Box+m^2})\f]\eqno(28)$$
with
$$W[\f,m^2,L,\L]=-{1\over2}\int_{\M
}[\f(\Box+m^2)\f]+{1\over2}Tr_{\L}{\rm ln}[{\Box+m^2\over
\L^2}]\eqno(29)$$
For simplicity we assume $\f$ and $m^2$ are  constant on ${\M}$,
consequently, treating $\L$ and $L$ as constants, $\cal G$ is a
2D mini-superspace. Keeping $\L$ finite ensures we have no UV
problems.

Examining the condition $dW=<dS>$ we obtain the equation
$$\int_{\M }[<(\Box+m^2)\phi>-(\Box+m^2)\f]d\f
-{1\over2}[\int_{\M}<(\phi-\f)^2>-Tr_{\L}({1\over\Box+m^2})]dm^2=0$$
This expression is finite without a cutoff only for $d=1$ where
$Tr({1\over\Box+m^2})$ converges, and corresponds to the
familiar situation of quantum mechanics. The metric induced on
this 2D space is
$$ds^2=\int_{\M }[m^2d\f^2]+{1\over
2}Tr_{\L}{1\over(\Box+m^2)^2}(dm^2)^2\eqno(30)$$
This metric does not need a cutoff to be well-defined for $d<4$,
however for $d=4$ $Tr({1\over\Box+m^2})^2$ is divergent and so
our metric is not well-defined without $\L$.

We can again look at the scalar curvature, which for the above
metric is of the form  $R={1\over4}Det^{-2}(g)\p_{m^2}Det(g)$.
Explicitly
$$R=-{Tr_{\L}(1+{\Box\over m^2})^{-3}\over {\left(Tr_{\L}(1+{\Box\over
m^2})^{-2}\right)}^2}
+{1\over2}{1\over Tr_{\L}(1+{\Box\over m^2})^{-2}}\eqno(31)$$
For $d=0$ this clearly reduces to the result for the Gaussian
distribution.
For ${\cal M}=(S^{1})^{d}$, $d=1,2$ or $3$, the cutoff can be
taken to zero giving in Fourier space
$$R=-{\sum_{n}{(1+({2\pi n\over mL})^2)}^{-3}\over
{\left(\sum_{n}{(1+({2\pi n\over mL})^2)}^{-2}\right)}^2}
+{1\over2}{1\over \sum_{n}{(1+({2\pi n\over
mL})^2)}^{-2}}\eqno(32)$$
In the limit $mL\r0$ the curvature reduces to the gaussian
curvature $R=-{1\over 2}$ while in the limit $mL\r\infty$ it
becomes
$$R=-{1\over 4}{(2-d)\over\Gamma({4-d\over2})}({m^2
L^2\over4\pi})^{-{d\over2}}+\dots\eqno(33)$$
This is a nice example of a crossover in the context of the
geometry of $\cal G$. For $d=2$, and $3$ the corrections in (33)
are exponentially small while for $d=1$ they are power law.
Interestingly for $2<d<4$ there is a crossover to positive
curvature which requires $R$ to pass through zero.

In a curved space setting for a conformally coupled free scalar
field
the formulae $(30)$ and $(31)$ for the metric and scalar
curvature remain unchanged. We can
consider a more general situation by including an additional
coupling to the curvature, this will be associated with the mass
term in a natural way. In the interacting case, treating $\l$ as
constant, i.e. we look only at the curvature of $\cal G$ in the
$\f$, $m^2$ plane, one would find that the curvature depended on
the scaling variable $Lm$ where $L$ is the characteristic length
scale of the geometry and $m$ is now the dressed mass of
equation (11) in this geometry. In the case of a totally finite
geometry the RG is of interest as physically there is a maximum
length scale in the problem. Hence the RG can only flow so far
before it stops. We also note that if we take a finite
temperature field theory [5] then $T={1\over L}$ and the
considerations of section 3 undergo a corresponding translation,
we therefore have a temperature dependent RG. In a real
cosmological setting one
can imagine including in a RG picture various other effects,
such as curvature, to get a quite detailed picture of  how the
universe cooled from the big bang. Naturally one would also wish
to generalize to the case of non-constant curvature where one
needs to consider a position dependent RG.
More discussion of these interesting matters in the context of
cosmology and the early universe will be discussed elsewhere
[5].

\section{Conclusions and Speculations} 

\def\T{\cal T}

The main aim of this paper has been to try to stimulate thought
along certain directions. There are certain problems that have
remained intractable for many years now: the confinement problem
in QCD and quantum gravity to name but two. We do not claim to
have solutions to problems such as these. We do claim, however,
that such theories exhibit certain key, common features, the
chief one being that the DOF in the problem are radically
different at different energy scales. We would also claim that
if this metamorphosis could be understood then a quantitative
understanding of the theory would probably follow.

The question of how systems behave under changes in scale is
most naturally addressed using the  field theoretic RG, a
consequence of an exact symmetry. However, there are, as pointed
out here, different, inequivalent representations of the RG. If
one has a field theory parametrized by a set of parameters
$P\equiv\{g^i\}$ corresponding to a point in ${\Gc}$ it might
occur that different subsets of the parameters, relevant for
describing the theory at  different scales, are taken into one
another by the RG flow on ${\Gc}$. If one's renormalization
depends only on a subset of the parameters one is restricting
one's flow to take place only in a subspace ${\T}$ of ${\Gc}$.
The resultant RG, $RG_{\T}$, depends only on a subset $K$ of the
parameters and the RG flows take place only on $\cal T$. If any
of the $P-K$ parameters are relevant in the RG sense then the
true RG flows of the theory, $RG_{\Gc}$, thought of as true
scale changes, will wish to flow off ${\T}$ into ${\Gc}$.
However, the use of $RG_{\T}$ does not allow for such flows.
Such a state of affairs would be shown up by the perturbative
unreliability of the results based on $RG_{\T}$. If none of the
parameters $K$ are relevant then there should be no problem.
However, one can only say what parameters are relevant when one
knows the full fixed point structure of the theory! In principle
it is obviously better to work with $RG_{\Gc}$. If a certain
parameter was important then one has made sure that its effects
are treated properly, and if it wasn't then that will come out
of the analysis. There can be no danger, except for extra work,
from keeping a parameter in, but there can be severe problems if
it is left out.
In the problems treated in this paper, although non-trivial,
they were easy in the sense that the parameter space ${\Gc}$ was
obvious. In the finite size case there were really 3 parameters
$m$, $\l$ and $L$. An $L$ independent RG was equivalent to
working on a 2 dimensional space which wasn't big enough to
capture the physics. What about QCD, or gravity? After all, in
QCD without fermions there appears to be only one parameter!
There is another length scale in QCD, the confinement scale,
however it is not manifest in the original Lagrangian, it comes
out dynamically. This length scale is the analog of $L$. As we
don't know how it really originates we arrive at a Catch 22
situation.  Our suggestion in such cases would be the following:
there are in most, if not all, of these type of problems
important classical field configurations; instantons, monopoles,
vortices etc. which are very important at one scale and not at
another. What one should do is derive a RG which is explicitly
dependent on such classical backgrounds just as we have shown in
section 4 that one should have a RG that is explicitly dependent
on one's background spacetime. This is contrary to the standard
view which tries to make a clean split between the background
(associated with IR effects) and renormalization of fluctuations
(which are usually taken to be associated with UV effects).
Although there may be scales where such an artificial split is
sensible it will certainly be true that there will be scales
where it manifestly is not.
We hope it is clear from the above that when we talk about a
parameter space it can be something quite complicated such as
that of the standard model, a very pertinent example of
crossover behaviour. We hope that we have convinced the reader
that there is a lot to be said for developing a RG that can
interpolate between different DOF.

We have considered here a class of problems that can be treated
so as to yield perturbatively the full crossover behaviour. In
section 4 we started to outline the most basic geometrical
elements of a more general framework for treating crossovers.
Our view was that a theory could be described by a set of
moments of a probability distribution that was a function of a
set of parameters. The idea was then to look at geometrical
structures on ${\Gc}$ to see: i) whether some non-perturbative
results could be gained in this way, and ii) whether through the
geometry one could obtain a better, geometrical understanding of
crossovers. It is obvious that in the more general setting we
are at a very rudimentary stage indeed. We do believe however,
that there are deep and important things to be learned from this
approach.

The geometry we looked at in section 4 was ordinary Riemannian
geometry based on a metric and a connection. There are many
questions to be asked. For instance, is the connection we
introduced the only relevant one? It would appear that
symplectic and contact geometry also play an important role.
There exists a symplectic form on the ``phase'' space composed
of the $\{g^i\}$ and their Legendre transform conjugates which
are expectation values of operators. There are also obvious
connections with the trace anomaly that we will not go into
here.
{}From a more physical point of view one would imagine the
intuition from lattice decimation could be extended to local
decimations which would lead to a position dependent RG. In this
setting the relevant geometry may be Weyl geometry. One may even
speculate [6] on cosmological expansion as a form of natural
decimation, where we are continuously decimating to scales
larger and larger than the Planck scale, or equivalently we are
following an RG flow further and further into the IR. One of the
problems in GR is the origin of time. There is from the
cosmological expansion of the universe a natural pinning of time
to energy scale, is this an accident? It may be that the
direction of time is due to gravity having an IR fixed point and
that we are only observing its RG flow as time.

{\bf Acknowlegements:} DOC is grateful to the Inst.~theor.~fys.,
Utrecht for travel support. CRS is grateful to FOM and DIAS for
financial support. We would like to thank variously Rafael
Sorkin, Gary Gibbons, Karel Kuchar and D.~V.~Shirkov for helpful
conversations.

\bibliographystyle{plain}

\end{document}